\documentclass[%
 reprint,
 amsmath,amssymb,
 aps,
]{revtex4-1}

\usepackage{graphicx}
\newcommand\sbullet[1][.5]{\mathbin{\vcenter{\hbox{\scalebox{#1}{$\bullet$}}}}}
\newcommand\scirc[1][.5]{\mathbin{\vcenter{\hbox{\scalebox{#1}{$\circ$}}}}}
\usepackage{graphicx, amsmath, amsfonts, tensor, amsthm, amssymb, bm, comment}
\usepackage{dcolumn}
\usepackage{bm, tikz-cd}

\def\TPNC{\text{teleparallelization}_{\text{NC}}}
\def\TPGR{\text{teleparallelization}_{\text{GR}}}

\begin{document}


\title{The Teleparallel Equivalent of Newton-Cartan Gravity}

\author{James Read}
\email{james.read@hertford.ox.ac.uk}
\affiliation{%
 Hertford College, University of Oxford \\
 Oxford OX2 6GG, United Kingdom}%
\author{Nicholas J.~Teh}
\email{nteh@nd.edu}
\affiliation{
University of Notre Dame \\ Notre Dame, IN 46556, USA
}%

\date{\today}

\begin{abstract}
We construct a notion of teleparallelization for Newton-Cartan theory, and show that the teleparallel equivalent of this theory is Newtonian gravity; furthermore, we show that this result is consistent with teleparallelization in general relativity, and can be obtained by null-reducing the teleparallel equivalent of a five-dimensional gravitational wave solution. This work thus strengthens substantially the connections between four theories:~Newton-Cartan theory, Newtonian gravitation theory, general relativity, and teleparallel gravity. 
\end{abstract}

\maketitle


\textit{Newton-Cartan theory and the teleparallel analogy}---Although Newton-Cartan theory (NCT) was originally conceived as a `geometrized' version of Newtonian gravitation \cite{Cartan}, this theory has recently been found to have a wide and impressive range of further applications:~it provides an improved model of the fractional quantum Hall effect \cite{GeracieSon, Son}, a geometric foundation for Horava-Lifschitz gravity \cite{Hartong}, and a framework for non-relativistic holography \cite{Christensen1, Christensen2}, to name just a few examples. 
Many of these applications owe their success to two powerful techniques for analyzing NCT. First, the recently-developed vielbein formalism for NCT \cite{GP, Andringa}  has played an indispensable role in understanding the hidden local Galilean invariance of NCT---akin to the hidden local Lorentz invariance of general relativity (GR)---and in analyzing the coupling of matter fields to a general Newton-Cartan background spacetime. Second, the technique of null dimensional reduction \cite{Nicolai, NR1}---which allows one to formulate a $D$-dimensional solution of NCT as a certain $(D+1)$-dimensional gravitational wave (Bargmann-Eisenhart) solution to GR---has provided an especially efficient method of using a relativistic spacetime to analyze the symmetries \cite{Duval} and dynamics \cite{Duval2} of a large class of non-relativistic mechanical systems.

Despite the obvious relevance and power of these techniques, they have not yet been applied to what is in some sense the \emph{fundamental theorem} of NCT, \emph{viz}.,~the Trautman Recovery Theorem \cite{Trautman}, which asserts that curved Newton-Cartan gravity is `empirically equivalent' to the standard (flat, but with forces) Newtonian theory of gravitation.
Furthermore, it seems to have gone unnoticed in the literature that such a statement is highly analogous to the key idea driving teleparallel gravity (TPG), which is that generically curved models of GR are empirically equivalent (modulo subtleties---see e.g.~\cite{Pereira, Nester}) to a theory with a flat but torsionful connection. 
This thus raises the question of why such an analogy exists, and whether one can use geometric tools to provide a satisfactory explanation of these relationships.
Indeed, on the basis of the analogy, one might suspect that standard Newtonian gravity is a kind of teleparallel theory, and---even more tantalizingly---that dimensional reduction of the kind described above `commutes' with this teleparallelization, in the sense that if we dimensionally reduce the teleparallel form of the Bargmann-Eisenhart solution, we precisely recover standard Newtonian gravity. This set of conjectural relationships can be illustrated schematically as follows:
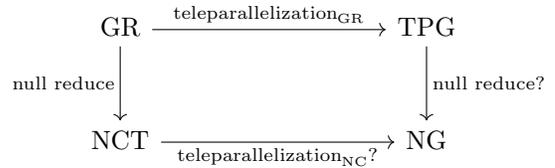
\begin{center}
\begin{tikzcd}[row sep=1cm, column sep=3cm]
\text{GR} \arrow[r, "\TPGR"] \arrow[d, "\text{null reduce}", swap]
& \text{TPG} \arrow[d, "\text{null reduce?}"] \\
\text{NCT} \arrow[r, "\TPNC ?", swap]
& \text{NG}
\end{tikzcd}
\end{center}
As will be discussed further below, $\TPGR$ is standardly defined in the TPG literature as the `gauging' of translations within the relativistic (Poincar\'{e}) gauge group while setting the spin connection to be pure gauge; its non-relativistic analog $\TPNC$ will be defined below. 

This letter answers the above question by combining vielbein and dimensional reduction techniques to shed new light on the Trautman Recovery Theorem, thereby showing that (i) Newtonian gravity is the teleparallel equivalent of Newton-Cartan theory (i.e.~by computing the bottom edge of the diagram), and (ii) that teleparallelization commutes with dimensional reduction (i.e.~by computing the top and right edges of the diagram).

While the relevance of the dimensional reduction technique to this result is obvious, it will perhaps help to highlight why the vielbein formulation of NCT plays such an important role in our argument. In the modern theory of TPG \cite{Pereira}, the vielbein apparatus is crucial for `gauging the translations' and thereby characterizing torsion as the field strength of this gauge field. Analogously, and as we will see below, if we want to teleparallelize NCT then we must first understand what the gauge group of NCT is from the perspective of its vielbein formulation---surprisingly, it turns out that `gauging the translations' in this context leads to the concept of a `mass torsion' that plays the role of force in Newtonian gravity. 

We proceed by first reviewing vielbein NCT and TPG for the reader. This done, we apply the teleparallelization techniques to vielbein NCT, demonstrating that the resulting theory is Newtonian gravitation theory, and that a notion of mass torsion is associated with force in this theory. This establishes the bottom leg of the above diagram. Finally, we construct the teleparallel equivalent of the Bargmann-Eisenhart solution of GR and show that it can be null reduced to obtain standard Newtonian gravitation, thereby completing the above diagram.

\textit{Review of vielbein NCT}---In this paper, we follow the index conventions of \cite{Berg, GP}---that is, for spacetime indices we use $M, N, \ldots$ running from 0 to 4; $\mu ,\lambda,\ldots$ running from 0 to 3; and the Latin index $v$ for a privileged spacetime null direction. On the other hand, for (internal) tangent space indices we use $I, J, \ldots$ running from 0 to 4; $a, b, \ldots$ running from 1 to 3; $A,B, \ldots$ running from 0 to 3; and $\pm$ for tangent space null directions (when in light-cone coordinates). We adopt a boldface notation when writing differential forms without indices---and switch freely between these notations as and when is convenient.

We now recall the vielbein formalism for NCT. Recent investigations (see e.g.~\cite[\S2.1]{GP}) have made it clear that the gauge group of 4D NCT is the Bargmann group (just as the gauge group of first-order---i.e. `vielbein + spin connection'---GR is the Poincar\'{e} group),
\begin{equation}
\mathrm{Barg} \left( 1, 3 \right) := \left( \mathrm{SO} \left(3\right) \ltimes \mathbb{R}^3 \right) \ltimes \left( \mathbb{R}^{3+1} \times {\mathrm{U}\left( 1\right)}_{\text{mass}} \right),
\end{equation}
where $\mathrm{Gal}_0 :=( \mathrm{SO} \left(3\right) \ltimes \mathbb{R}^3 )$ is the homogeneous Galilean group containing rotations and boosts, $\mathbb{R}^{3+1}$ represents space and time translations, and the central extension $\mathrm{U}(1)_{\text{mass}}$ represents translations along a `mass dimension'. 
The corresponding Bargmann algebra has the non-zero commutation relations
\begin{align}
\left[ J_{ab} , J_{cd} \right] &= 4 \delta_{[a[c} J_{d]b]} , \\
\left[ J_{ab} , P_c \right] &= -2 \delta_{c[a}P_{b]} , \\
\left[ J_{ab} , G_c \right] &= -2 \delta_{c[a} G_{b]} , \\
\left[ G_a, H \right] &= -P_a , \\
\left[ G_a , P_b \right] &= -\delta_{ab} Z,
\end{align}
where $Z$ is a central charge  generating translations along an internal `mass' dimension, and $H$, $P_a$, $G_a$, and $J_{ab}$ generate time translations, spatial translations, Galilean boosts, and spatial rotations, respectively.

Recall that in GR, a vielbein $e\indices{_{\mu}^{A}}$ is a geometrical object which implements a change-of-basis so as to diagonalise the metric field---we have $g_{\mu\lambda} = e\indices{_{\mu}^{A}} e\indices{_{\lambda}^{B}} \eta_{AB}$. Since vielbeins are valued in the quotient algebra $\mathbb{R}^{1,d} \cong \mathfrak{poinc}/\mathfrak{lor}$, they are associated with the gauge group of translations.
By analogy with this familar construction, an extended vielbein \cite{GP} for NCT theory is a $1$-form field $e\indices{_{\mu}^{I}} = ( \tau_\mu, e\indices{_{\mu}^{a}}, m_{\mu} )$, which is valued in the extended internal tangent space $\mathbb{R} \oplus \mathbb{R}^3 \oplus {\mathfrak{u}\left( 1\right)}_{\text{mass}}$ (i.e.,~the direct sum of infinitesimal time translations, spatial translations, and mass translations). Following \cite{GP}, we equip the extended internal tangent space with a Minkowski metric $\eta_{IJ}$, and use light-cone coordinates in which $\eta_{ab}=\delta_{ab}$ and $\eta_{-+}=\eta_{+-}=1$. As extensively discussed in \cite{GP}, this vielbein is `extended' in the sense that it includes the $\mathrm{U}(1)_{\text{mass}}$ gauge field $m_\mu$, which is needed to correctly couple a massive point particle (and matter fields) to a Newton-Cartan background. We also note that a (non-extended) frame $e\indices{^{\mu}_{A}} = \left(v^\mu, e\indices{^{\mu}_{a}}\right)$ can be defined in the usual way by means of $e\indices{^{\mu}_{A}} e\indices{^{A}_{\lambda}} = \delta\indices{^{\mu}_{\lambda}}$ and $e\indices{_{\mu}^{A}} e\indices{^{\mu}_{B}} = \delta\indices{^{A}_{B}}$. 

The associated (extended) spin connection $1$-form $\omega\indices{_{\mu}^{I}_{J}}$ is valued in $\mathfrak{gal}_0$ and consists of the boost connection $\omega\indices{_{\mu}^{a}_{b}}$ and the rotation connection $\omega\indices{_{\mu}^{a}}:= \omega\indices{_{\mu}^{a}_{0}}$ (below, we will discuss how the spin connection is related to  a familiar spacetime connection). 
We note that $e\indices{_{\mu}^{I}}$ and $\omega\indices{_{\mu}^{I}_{J}}$ transform under local Galilean boosts of the form $v^\mu \mapsto v^\mu + k^b e\indices{_{b}^{\mu}}$.

The above objects constitute the \textit{Cartan data} for locally defining a Cartan gauge theory (for further details, see e.g.~\cite{Andringa, GP}). The gauge covariant derivative is defined as $D {\bm{\alpha}}^I := d {\bm{\alpha}}^I + {\bm{\omega}}\indices{^{I}_{J}} \wedge {\bm{\alpha}}^J$, which in turns allows one to define the Cartan torsion $\bm{T}^I = D\bm{e}^I = d\bm{e}^{I} + \bm{\omega}\indices{^{I}_{J}} \wedge \bm{e}^J$ and the Cartan curvature $\bm{R}\indices{^{I}_{J}} = d\bm{\omega}\indices{^{I}_{J}} + \bm{\omega}\indices{^{I}_{K}}\wedge \bm{\omega}\indices{^{K}_{J}}$. We now label these internal torsions and curvatures by their respective generators (cf.~\cite{Andringa}): 

\begin{align}
(\bm{f})_{\mu \lambda}:=T_{\mu\lambda}\left(M\right) &= 2\partial_{[\mu} m_{\lambda]} - 2 \omega\indices{_{[\mu}^{a}}e_{\lambda]a}, \\
T_{\mu\lambda}\left( H \right) &= 2 \partial_{[\mu} \tau_{\lambda]} , \\
T\indices{_{\mu\lambda}^{a}} \left( P\right) &= 2 \partial_{[\mu} e\indices{_{\lambda]}^{a}} - 2 \omega\indices{_{[\mu}^{ab}} e_{\lambda]b} - 2\omega\indices{_{[\mu}^{a}} \tau_{\lambda]} , \\
R\indices{_{\mu\lambda}^{a}}\left( G \right) &= 2 \partial_{[\mu} \omega\indices{_{\lambda]}^{a}} - 2\omega\indices{_{[\mu}^{ab}}\omega_{\lambda]b} , \\
R\indices{_{\mu\lambda}^{ab}} \left(J\right) &= 2 \partial_{[\mu} \omega\indices{_{\lambda]}^{ab}}.
\end{align}
Since we are interested in Newtonian gravity, we will only consider cases of NCT in which $T_{\mu\lambda}\left(H\right)=T\indices{_{\mu\lambda}^{a}}\left(P\right)=0$ (cf.~\cite{Andringa, Berg}). We further stress that the internal Cartan mass torsion $\bm{f}$ cannot be converted into a spacetime torsion because the extended vielbein is not invertible. 


A Newton-Cartan (NC) spacetime is defined as $(M, h^{\mu \lambda}, \tau_\mu)$, where $M$ is a smooth $4$-manifold, $h^{\mu \lambda}:= \delta^{ab} e\indices{^{\mu}_{a}} e\indices{^{\lambda}_{b}}$ is a degenerate `spatial metric' (which will be used to raise indices), and the clock 1-form $\tau_
\mu$ plays the role of a `temporal metric' (thus, the orthogonality condition $h^{\mu \lambda} \tau_\lambda = 0$ holds because  $e\indices{^{\mu}_{a}}$ is spacelike). Although $h^{\mu \lambda}$ is non-invertible, one can still construct the manifestly frame-dependent covariant tensor $h_{\mu \lambda}:= \delta_{ab}e\indices{^{a}_{\mu}} e\indices{^{b}_{\lambda}}$, which transforms under local Galilean boosts acting on frames.
To discuss such frame-dependent quantities, it will be convenient to introduce the notion of a \textit{observer vector field}, \emph{viz}.,~a vector field $n^\mu$ that is time-like and normalized ($\tau_\mu n^\mu = 1$).

Following standard procedure, we can use the vielbein and the spin connection to equip NC spacetime with a compatible spacetime connection $\widetilde{\nabla}$ (i.e.,~$\widetilde{\nabla}_\mu h^{\sigma \lambda} = \widetilde{\nabla}_\mu \tau_\lambda = 0$), by means of the vielbein hypothesis $\widetilde{\nabla}_\mu e\indices{_{\lambda}^{A}} + \omega\indices{_{\mu}^{A}_{B}} e\indices{_{\lambda}^{B}}=0$. 
Explicitly, $\widetilde{\nabla}$ has the Christoffel symbols
\begin{equation}\label{NCGamma}
    \widetilde{\Gamma}\indices{^\lambda_{\mu \sigma}} = \overset{n}{\Gamma}{}\indices{^{\lambda}_{\mu \sigma}} + \tau_{\mu}F\indices{_\sigma ^\lambda},
    \end{equation}
where $n^\lambda$ is an observer vector field,  $\overset{n}{\Gamma}{}\indices{^{\lambda}_{\mu \sigma}} := n^\lambda \partial_{(\mu}\tau_{\sigma)} + \frac{1}{2}h^{\lambda \rho} (\partial_{\mu} h_{\sigma \rho} + \partial_{\sigma} h_{\mu \rho}- \partial_{\rho} h_{\mu \sigma})$, and $(\bm{F})_{\sigma \lambda} =(\bm{\omega}_a \wedge \bm{e}^a)_{\sigma \lambda}$ is a 2-form. 

(\ref{NCGamma}) shows us that unlike Lorentzian spacetime, NC spacetime does not induce a unique compatible torsion-free spacetime connnection. Instead, the space of possible connections can be parametrized by fixing an observer field $n^\lambda$ which determines $\overset{n}{\Gamma}$, and then further specifying $F\indices{_\sigma ^\lambda}$ in order to pick out $\widetilde{\Gamma}\indices{^\lambda_{\mu \sigma}}$. 
These two pieces of data have an important physical interpretation:~$\overset{n}{\Gamma}$ is an `inertial connection' in the sense that $n$ is acceleration-free and vorticity-free with respect to $\overset{n}{\Gamma}$, and $F\indices{_\mu ^\sigma} = \tau_\mu \alpha^\sigma + \gamma\indices{_\mu ^\sigma}$, where $\alpha^\mu := n^\sigma \widetilde{\nabla}_\sigma n^\mu $ is the \textit{acceleration} and $\gamma_{\mu\kappa} := e\indices{^a_{[\mu}} e\indices{^b_{\kappa]}} e\indices{^{\lambda}_{a}} e\indices{_{\sigma b}} \widetilde{\nabla}_\lambda n^\sigma$ is the \textit{vorticity} of $n^\mu$ with respect to $\widetilde{\Gamma}$.
Finally, we will find it useful to note that the spacetime boost connection $\omega\indices{^{\mu}_{\lambda}} := \omega\indices{_{\mu}^{a}}e\indices{_{a}^{\lambda}}$ can be written \cite{GP}
\begin{equation}\label{boostconn}
    \omega\indices{_{\mu}^{\lambda}} = \theta\indices{_{\mu}^{\lambda}} + \tau_\mu \alpha^\lambda + \gamma\indices{_{\mu}^{\lambda}},
\end{equation}    
where $\theta_{\mu\kappa} :=
e\indices{^{a}_{(\mu}} e\indices{^{b}_{\kappa)}}
 e\indices{^{\lambda}_{a}} e_{\sigma b} \widetilde{\nabla}_\lambda n^\sigma$ is the \textit{expansion} of $n^\mu$.

In order to define classical NCT, the data $(M, e\indices{_{\mu}^{I}}, \omega\indices{_{\mu}^{I}_{J}})$ must be supplemented with a mass density scalar field $\rho$ and an observer field $\xi^\mu$ that represents test particle trajectories, as well as the `Newtonian condition' $d\bm{F}=0$ and the vanishing rotational curvature condition $R\indices{_{\mu\lambda}^{ab}}\left(J\right)=0$.
We say that this data is an \textit{NCT model} if it satisfies the geodesic equation $ \xi^\mu \widetilde{\nabla}_\mu  \xi^\lambda =0$
and the source equation \begin{equation}\label{NCTEOM} \widetilde{R}_{\mu \lambda} = 4 \pi \rho \tau_\mu \tau_\lambda.\end{equation}

On the other hand, the data of standard Newtonian gravitation (NG) is given by $(M, h^{\mu\lambda}, \tau_{\mu}, \nabla, \rho, \xi^\mu, \phi)$, where $\nabla$ is flat and compatible, and $\phi$ is a scalar gravitational potential; we say that this data is an \textit{NG model} if it satisfies the force equation $\xi^\mu \nabla_\mu  \xi^\lambda = - \nabla^\lambda \phi$ and the Poisson equation $\nabla_\mu \nabla^\mu \phi = 4 \pi \rho$. We can now state the central foundational result of NCT, \emph{viz}.,~the Trautman Recovery Theorem \cite{Trautman}:~given an NCT model, one can locally define an empirically equivalent NG model up to the shift 
\begin{equation}\label{NGshift}
\left(\Gamma\indices{^{\lambda}_{\mu\sigma}}, \phi \right) \mapsto \left(\Gamma\indices{^{\lambda}_{\mu\sigma}} + \tau_\mu \tau_\sigma \nabla^\lambda \psi,~~ \phi + \psi \right),
\end{equation}
where $\nabla^\mu \nabla^\sigma \psi = 0$. (For a detailed discussion of this theorem and its implications, see \cite{Malament, Teh}.)

\textit{Teleparallel equivalent of GR (TPG)}---We now review the modern approach to TPG \cite{Pereira}. 
Vielbein GR and TPG both use the Cartan data of an $\mathbb{R}^{1,d}$-valued coframe $e\indices{_{M}^{I}}$ and a Lorentz spin connection $\omega\indices{_{M}^{I}_{J}}$, albeit in different ways.
In vielbein GR, one starts with the inertial degrees of freedom of special relativity (SR) and generalizes to GR by `gauging' the spin connection (i.e.,~treating it as a source of Cartan curvature), thus arriving at the Levi-Civita spacetime connection $\overset{\scirc}{\nabla}$ which captures both gravitational and inertial degrees of freedom. By contrast, TPG restricts the spin connection to only represent the inertial effects present in a particular coframe and generalizes SR by `gauging' the coframe to obtain torsion (which, as we mentioned above, is just the field strength of the coframe).
More precisely, TPG requires the spin connection to take the schematic form $\omega = \Lambda \partial \Lambda$ (where $\Lambda$ is a local Lorentz transformation), which is the most general condition under which the Cartan curvature vanishes. 
In practice, it is often convenient to gauge-fix to the case where $\omega\indices{_{M}^{I}_{J}}=0$, thus restricting to the class of frames (`proper frames') in which inertial effects are absent. One can then use the vielbein hypothesis to define the (unique) associated flat Weitzenb\"{o}ck spacetime connection $\overset{\sbullet}{\nabla}$.
We recall that the relationship between $\overset{\scirc}{\nabla}$ of GR and $\overset{\sbullet}{\nabla}$ of TPG is given by
\begin{equation}\label{GGK}
\overset{\sbullet}{\Gamma}{}\indices{^{M}_{NP}} = \overset{\scirc}{\Gamma}{}\indices{^{M}_{NP}} + K\indices{^{M}_{NP}},
\end{equation}
\noindent where $K\indices{^{M}_{NP}}$ is the \textit{contorsion tensor}.

Vacuum TPG is defined by the data $\left(M, e\indices{_{M}^{I}}\right)$, and solutions of the theory satisfy the equation of motion
\begin{equation}\label{j}
\partial_N \left( e S\indices{_I^{NM}} \right) - 4\pi \left( e j\indices{_{I}^{M}} \right) = 0 ,
\end{equation}
\noindent where $e := \det\left( e\indices{_{M}^{I}} \right) $ and the teleparallel gravitational current, superpotential, and Lagrangian are respectively given by:
\begin{align*}
e j\indices{_{I}^{M}} &= -\frac{1}{4\pi} e e\indices{_{I}^{N}}S\indices{_{P}^{QM}}T\indices{^{P}_{QN}} + e\indices{_{I}^{M}}\mathcal{L}_G, \\
S^{MNP} &= \frac{1}{2} \left( K^{NPM} - g^{MP}T\indices{^{QN}_{Q}} + g^{MN}T\indices{^{QP}_{Q}} \right), \\
\mathcal{L}_G &= \frac{e}{16\pi} S^{MNP}T_{MNP},
\end{align*}
where $T\indices{^{M}_{NP}}:=  \overset{\sbullet}{\Gamma}{}\indices{^{M}_{NP}}-\overset{\sbullet}{\Gamma}{}\indices{^{M}_{PN}}$ is the \textit{torsion tensor}, which encodes the antisymmetric part of the connection.

A straightforward calculation shows that \eqref{j} is equivalent to the vacuum Einstein equation $R_{MN}=0$. $\mathcal{L}_G$ is equivalent to the Einstein-Hilbert Lagrangian up to a boundary term---cf.~\cite[p.~92]{Hehl}. And under mild assumptions (in particular, covariant conservation of the source stress-energy tensor which would appear on the right hand side of \eqref{j} in the non-vacuum case), 
it follows from application of the Mathisson-Papapetrou method for deriving particle equations of motion \cite{LeclercMP} (cf.~also \cite{GJ}) that test particles satisfy the teleparallel force equation
\begin{equation}\label{TPGforce}
\xi^M \overset{\sbullet}{\nabla}_M \xi_N = \xi^P T_{PNM} \xi^M .
\end{equation}


Clearly, there exists a striking analogy between the flat NG connection $\nabla$ in the Trautman recovery theorem and the flat Weitzenbock connection $\overset{\sbullet}{\nabla}$ obtained through $\TPGR$. We now demonstrate that it is not only possible to understand Trautman's result in terms of an analogous $\TPNC$, but that, remarkably, this procedure commutes with $\TPGR$ via null reduction.

\textit{Newton-Cartan teleparallelization}---In keeping with the teleparallel philosophy, we separate out the gravitational and inertial degrees of freedom of NCT, by locating the former in the `mass torsion' and the latter in a suitably gauge-fixed spin connection.
As above, tildes denote the NCT objects, whereas we omit any oversetting for the associated teleparallel objects.

By analogy with the torsionlessness of GR, we require that the mass torsion of NCT vanish, i.e.~$\bm{\tilde{f}}=0$. 
From (7), we thus see that the boost curvature is related to the mass gauge field by $d\bm{\tilde{m}} = \bm{\tilde{\omega}}_a \wedge \bm{\tilde{e}}^a$. Note that the constraints $\bm{\tilde{f}}=\tilde{T}\indices{_{\mu\lambda}}\left(H\right)=\tilde{T}\indices{_{\mu\lambda}^{a}}\left(P\right)=0$ allow one to determine $\bm{\tilde{\omega}}^I$ solely in terms of $\bm{\tilde{e}}^I$, thus determining the NCT spacetime connection $\widetilde{\nabla}$.

We now introduce teleparallel NCT by identifying the Cartan data $(e\indices{_{M}^{I}}, \omega\indices{_{M}^{I}_{J}})$ that yields (i) a flat spacetime connection $\nabla$ (the analog of the Weitzenbock connection of TPG), and (ii) whose Cartan torsion compensates for the curvature of $\widetilde{\nabla}$ (the analog of the Levi-Civita connection of GR). While one might be tempted to introduce non-trivial spatial torsion, we note that for a flat connection, the Bianchi identities imply that $T\indices{_{\mu\lambda}^{a}}\left(P\right)=0$, so it is only the mass torsion $\bm{f}$ that is relevant here. 

First, we perform a preliminary `inertial gauge-fixing' that is motivated by \eqref{NCGamma} and the discussion following it: recall that $(e\indices{_{M}^{I}},\omega\indices{_{M}^{I}_{J}})$ determines the inertial connection $\overset{v}{\Gamma}{}\indices{^{\lambda}_{\mu \sigma}}$ precisely when it is subject to the constraint $\bm{F} =0$, upon which $\bm{f}=d\bm{m}$. Implementing this constraint and setting $\bm{\tilde{m}}=\bm{m}$ then yields the following relationship between the NC and teleparallel equations of motion:
\begin{equation}\label{geodesicforce1}
    \xi^\mu \widetilde{\nabla}_\mu \xi^\lambda = 0 \quad \Longleftrightarrow \quad \xi^\mu \nabla_\mu \xi^\lambda = f\indices{^{\lambda}_{\mu}} \xi^\mu.
\end{equation}
Evidently, the tensor $\tau_\mu F\indices{_{\sigma}^{\lambda}}$ is the analogue of the contorsion in this scenario, and it is determined by the teleparallel mass torsion $\bm{f}$.


Clearly, the above gauge-fixing is not invariant under local Galilean boosts.
We can remedy this by transforming a vielbein to the `twistless gauge' by means of $\bm{e}^a \mapsto \bm{\bar{e}}^a= \bm{e}^a - m^a \bm{\tau}$, where $m^a = e^{\mu a} m_\mu$, thus resulting in the new extended vielbein $\bm{\bar{e}}^I= \left( \bm{\tau} ,\bm{\bar{e}}^a, \bm{\tau} \phi \right) $, where $\phi$ is a boost-invariant scalar. The corresponding frame contains a boost-invariant observer field $z^\mu := \bar{e}_{0}^\mu =e_{0}^{\mu} - h^{\mu \lambda} m_\lambda $. 
This choice is invariant under local boosts acting on the unbarred quantities; however, there is still a residual boost-freedom that is parameterized by the $\mathrm{U}(1)_{\text{mass}}$ gauge transformations $\bm{m} \mapsto \bm{m} + df$ (where $f$ is an arbitrary smooth function), under which $(z^\mu, \phi)$ is transformed into another boost-invariant pair $(z'^\mu, \phi')$.

The name `twistless gauge' stems from the observation that the vorticity of $z^\mu$ vanishes with respect to $\widetilde{\nabla}$, so our frame is twistless with respect to the NCT connection. Since $\bm{f}= \bm{\tau} \wedge \bm{d\phi}$ in this gauge, we see from \eqref{geodesicforce1} that
\begin{equation}
  \xi^\mu \widetilde{\nabla}_\mu \xi^\lambda = 0 \quad \Longleftrightarrow \quad \xi^\mu \nabla_\mu \xi^\lambda = - h^{\lambda \mu} (d\phi)_\mu,
\end{equation}
thus recovering the NG force law as the teleparallel equation of motion.

Having gauge-fixed to the inertial connection $\overset{z}{\Gamma}{}\indices{^{\lambda}_{\mu \sigma}}$, we see from \eqref{boostconn} that there is only one parameter left to gauge-fix in $\omega\indices{_{\mu}^{\lambda}}$, \emph{viz}.~the expansion $\theta\indices{_{\mu}^{\lambda}}$. We do so by setting $\omega\indices{_{\mu}^{\lambda}}= \theta\indices{_{\mu}^{\lambda}} = 0$, which along with $R\indices{_{\mu\lambda}^{ab}}\left(J\right)=0$ (cf.~the definition of an NCT model) implies that $\overset{z}{\Gamma}{}\indices{^{\lambda}_{\mu \sigma}}$ is flat. One can then easily deduce that the NCT source equation holds just in case the flat teleparallel connection $\overset{z}{\nabla}$ satifies the NG Poisson equation. 

We have absorbed the Trautman recovery theorem into $\TPNC$, but for the shift symmetry \eqref{NGshift}. In fact, this symmetry is accounted for by the fact that the data $(\bm{\bar{e}}^I,\bm{\omega}\indices{^{I}_{J}})$ satisfying the twistless gauge and $\omega\indices{_{\mu}^{\lambda}}=0$ is unique only up to $\mathrm{U}(1)_{\text{mass}}$ gauge transformations that preserve these conditions. 

\emph{Null reduction commutes with teleparallelization}---Having defined $\TPNC$, we show now that, when intertwined with null reduction, this procedure commutes with $\TPGR$. In so doing, we also give the first explicit teleparallel description of the Bargmann-Eisenhart (BE) solution. 

The BE solution \cite{Berg, NR1} is a 5D vacuum solution of GR equipped with a null vector field $\Xi$ that is parallelized by the Levi-Civita connection (and is thus Killing). It is well-known that 4D NC spacetime can be constructed as the orbit space of the flow generated by $\Xi$, which is the so-called `null reduction'. Since comprehensive discussions of this familiar technique have already been given in \cite{Nicolai, NR1}, here we will only introduce the form that is best-suited to our computations, \emph{viz}.~the Julia-Nicolai ansatz \cite{Nicolai} for the BE solution, which directly incorporates null reduction by parameterizing the 5D relativistic frame field $\hat{e}\indices{^{M}_{I}}$ in terms of 4D NC fields as follows: the only non-zero components of $\hat{e}\indices{^{M}_{I}}$ are $\hat{e}\indices{^{\mu}_{a}} = e\indices{^{\mu}_{a}}$, $\hat{e}\indices{^{v}_{a}} = e\indices{^{\mu}_{a}} m_\mu$, $\hat{e}\indices{^{\mu}_{-}} = v^\mu$, $\hat{e}\indices{^{v}_{-}} = v^\mu m_\mu$, and $\hat{e}\indices{^{v}_{+}} =1$, where the $v$ spacetime-direction is aligned with $\Xi$. The corresponding co-frame field $\hat{e}\indices{_{M}^{I}}$ has the non-zero components $\hat{e}\indices{_{\mu}^{a}} = e\indices{_{\mu}^{a}}$, $\hat{e}\indices{_{\mu}^{-}} = \tau_\mu$,
$\hat{e}\indices{_{\mu}^{+}} = -m_\mu$, and
$\hat{e}\indices{_{v}^{+}} = 1$. (Note that as compared with \cite[\S III]{Berg}, we set the compensating field $S=1$.)
We again stress that the power of this ansatz stems from viewing the BE solution as an $\mathbb{R}$-bundle over $M$ (with fibers generated by $\Xi$) and making a judicious choice of section that lets us express the 5D fields entirely in terms of the 4D NC fields contained in $e\indices{_{\mu}^{I}}$; thus the null reduction can simply be read off the $\mu$-components of the coframe field.
Furthermore, at least locally, we can use a diffeomorphism to transform the ansatz into the `twistless gauge' where $\bm{m} = \bm{\tau} \phi$ and we will assume this gauge-fixing henceforth (cf.~\cite{Nicolai}); this is of course consistent with the fact that the null reduction of any BE solution automatically satisfies the Newtonian condition, as shown in \cite{NR1}.
We note that \cite{Berg2} showed that if the 5D GR fields are on-shell, their null reduced NC fields will have $T_{\mu\lambda}\left(H\right)=T\indices{_{\mu\lambda}^{a}}\left(P\right)=0$.
Since we will now construct the pure tetrad TPG version of the BE solution, the spin connection vanishes and so this condition amounts to $d\bm{\tau}=d\bm{e}^a=0$.

Given the above ansatz, the two non-zero components of the 5D Weitzenb\"{o}ck connection read
\begin{align}
\Gamma\indices{^{\mu}_{\kappa\lambda}} &= e\indices{^{\mu}_{a}} \partial\indices{_{(\kappa}} e\indices{^{a}_{\lambda)}} + z^\mu \partial\indices{_{(\kappa}} \tau\indices{_{\lambda)}} , \label{G1} \\
\Gamma\indices{^{v}_{\kappa\lambda}} &= e\indices{^{\sigma}_{a}}m_\sigma \partial_\kappa e\indices{^{a}_{\lambda}} - \partial_\kappa m_\lambda + z^\sigma m_\sigma \partial_\kappa \tau_\lambda .
\end{align}
\noindent (Here, we have symmetrized indices in \eqref{G1}, since the torsion associated with this connection component vanishes on-shell.) From this, one finds that all components of the torsion tensor vanish, save
\begin{equation}
T\indices{^{v}_{\kappa\lambda}} = - 2 \partial_{[\kappa} m_{\lambda]} .
\end{equation}
The non-vanishing components of contorsion are ($z^\mu$ is the boost-invariant observer field, defined above)
\begin{align}
K\indices{^{\mu}_{\kappa\lambda}} &= -\tau_\kappa h^{\mu\beta} \partial_{[\beta} m_{\lambda]} - \tau_\lambda h^{\mu\beta} \partial_{[\beta} m_{\kappa]},\label{K1} \\
K\indices{^{v}_{\kappa\lambda}} &= -\tau_\kappa z^\beta \partial_{[\beta}m_{\lambda]} - \tau_\lambda z^\beta \partial_{[\beta}m_{\kappa]} - 2\partial_{[\kappa} m_{\lambda]} . \label{K2}
\end{align}
(Some of these components are related to anholonomity---cf.~\cite[p.~39]{Hehl}.) From these components of contorsion, one can compute the components of the superpotential $S^{MNP}$, and thereby check that $\mathcal{L}_G=0$ as expected, since we assumed at the outset that the theory is on-shell.

To show that this teleparallelized theory just is NG, it remains to derive the NG force equation for test particles, and the Poisson equation.
For the former, recall that the only non-vanishing component of torsion is $T\indices{^{v}_{\sigma\lambda}}$ and so the TPG force equation \eqref{TPGforce} yields
\begin{equation}
\xi^M \overset{\sbullet}{\nabla}_M \xi_\sigma = \xi_v T\indices{^{v}_{\sigma\mu}}\xi^\mu 
= -2 \xi_v \partial_{[\sigma} m_{\mu]} \xi^\mu .
\end{equation}
Using the gauge-fixing condition $m_\mu = \tau_\mu \phi$ and metric compatibility, we have 
\begin{equation}
\xi^\mu \overset{\sbullet}{\nabla}_\mu \xi^\sigma = - \overset{\sbullet}{\nabla}{}\indices{^\sigma} \phi,
\end{equation}
which is the Newtonian force equation.

To obtain the Poisson equation, we first note that we can write \eqref{NCTEOM} in terms of the boost curvature $R\indices{_{\mu\lambda}^{a}}\left(G\right)$, which in turn implies that $R_{--}=4\pi\rho$, where $R_{IJ}= \hat{e}\indices{^M_K} \hat{e}\indices{^N_I}R\indices{_{MN}^K_J}$. 
If we then impose the $R\indices{_{\mu\lambda}^{ab}}\left(J\right)=0$ condition, then it is easy to see (cf.~\cite[\S3]{Berg}) that the only non-zero component of $R_{IJ}$ is $R_{--}$; using $\overset{\scirc}{R}_{MN} = \hat{e}\indices{_{M}^{I}} \hat{e}\indices{_{N}^{J}} R_{IJ}$, we thus have
\begin{equation}
\overset{\scirc}{R}_{\mu\lambda} = \tau_\mu \tau_\lambda R_{--}
= 4 \pi \rho \tau_\mu \tau_\lambda \label{P1}.
\end{equation}

To relate this quantity to the teleparallel connection, we recall that the Levi-Civita Ricci tensor can be written in terms of contorsion as (cf.~\cite[p.~3]{TPG1})
\begin{multline}
\overset{\scirc}{R}_{NQ} = \overset{\scirc}{\nabla}_{Q} K\indices{^{M}_{NM}} - \overset{\scirc}{\nabla}_{M} K\indices{^{M}_{NQ}} + \\ K\indices{^{R}_{NM}}K\indices{^{M}_{RQ}} - K\indices{^{R}_{NQ}}K\indices{^{M}_{RM}}.
\end{multline}
Using \eqref{GGK}, \eqref{K1}, \eqref{K2}, and $m_\mu = \tau_\mu \phi$, one finds for $\overset{\scirc}{R}_{\mu\sigma}$ that
\begin{equation}
    \overset{\scirc}{R}_{\mu\sigma} = - \overset{\sbullet}{\nabla}_{\lambda} K\indices{^{\lambda}_{\mu \sigma}} 
    = \overset{\sbullet}{\nabla}_{\lambda} (\tau_\mu \tau_\sigma h^{\lambda \beta} d\phi_{\beta}) . \label{P2}
\end{equation}
Combining \eqref{P1} and \eqref{P2} and using compatibility, we recover precisely the Poisson equation of NG,
\begin{equation}\label{poisson}
h^{\mu\lambda} \overset{\sbullet}{\nabla}_\mu \overset{\sbullet}{\nabla}_\lambda \phi = 4\pi \rho.
\end{equation}

Finally, recent investigations have shown that TPG contains a gauge symmetry that can be expressed covariantly as the `weak gauge transformations' of \cite{BW}, or from the Hamiltonian perspective as the $\lambda$-symmetry introduced in \cite{Blag}. 
When these TPG gauge transformations preserve the above gauge-fixing, they give rise to the shift symmetry \eqref{NGshift} of NG.



\textit{Outlook.}---The foregoing both offers us deeper insight into the foundations of Newtonian gravitation, and opens up a rich array of possible future research directions. For instance, it is well-known that NG faces problems of consistency in infinite, homogeneous, isotropic universes (see e.g.~\cite[\S4.4]{Malament} and references therein). Roughly, one needs to pick a privileged centre point in any given solution---but then the magnitude of the gravitaitonal force is arbitrary. 
The now-standard solution to this problem is the following:~different solutions correspond to different gauge-fixings (which define NG) of a particular curved derivative operator of NCT, and consistency can be restored by moving to the fully gauge-invariant picture offered by NCT.

We observe that an analogous consistency issue seems to arise in the case of a historic (but mistaken!) attempt at defining TPG, i.e.~the so-called `pure tetrad' version of TPG (see \cite{kop} for the original source, and \cite{Leclerc} for contemporary discussion). Unlike modern TPG which is fully gauge-invariant  and consistent \cite{Pereira} (though see \cite{Maluf1, Maluf2} for some concerns regarding the Hamiltonian formulation of covariant TPG, and \cite{Blag2, Ferraro} for replies), the pure tetrad version privileges a particular `proper frame'; our work shows that the inconsistency of NG is really a special case of what happens when one does this. To see the analogy explicitly, suppose we gauge-fix the spin connection to vanish. Given this, we can determine a unique $g_{\mu\lambda}$; however, the tetrad field $e\indices{^{a}_{\mu}}$ is determined only up to a Lorentz transformation, $e\indices{^{a}_{\mu}} \mapsto \Lambda\indices{^{a}_{b}}e\indices{^{b}_{\mu}}$ (here, Latin indices are four-dimensional tangent space indices). This is unproblematic for spinless matter, which couples to the metric alone. However, in the pure tetrad theory, matter fields with spin couple to the tetrad directly---so their behaviour will depend upon a choice of frame. Thus, just as in the case of NG, we have an unpalatable gauge-dependence of the `force' theory which may be resolved on moving to the associated `geometrized' or `covariantized' theory.

Second, many applications of NCT to condensed matter physics (e.g.~\cite{GeracieSon}) use a generalization of NCT called \emph{twistless-torsional NCT}, which allows some degree of temporal torsion but still requires that the solutions remain causal (in the sense that $\bm{\tau} \wedge d\bm{\tau} = 0$, which ensures the existence of hypersurfaces of absolute simultaneity). 
Furthermore, it was shown in \cite{Berg} that twistless-torsional NCT can be constructed via the null reduction of (on-shell) five-dimensional conformal gravity.
Since there exist well-known conformal generalizations of TPG (e.g.~\cite{Maluf, Bamba2013}), our work paves the way for applying conformal TPG to condensed matter physics via a null reduction. In particular:~can one construct a diagram, relating conformal gravity, conformal TPG, twistless-torsional NCT, and the teleparallelization of twistless-torsional NCT, in exactly the manner in which the diagram relating GR, TPG, NCT, and NG was constructed in this paper?  
We expect that doing so will be particularly useful in applications in which it is helpful to make a judicious choice of inertial frame, as is often the case when analyzing the behavior of fluids and lattices (cf.~\cite{GeracieSon}). 

\section*{Acknowledgements} J.R.~is supported by an AHRC scholarship at the University of Oxford, and a predoctoral fellowship at the University of Illinois at Chicago, under grant number 56314 from the John Templeton Foundation. N.T.'s work on this project was partially funded by NSF Award 1734155.



\bibliographystyle{apsrev4-1}

\end{document}